\documentclass[12pt]{article}
\usepackage{epsfig}
\def\be{\begin{equation}}
\def\ee{\end{equation}}
\def\bea{\begin{eqnarray}}
\def\eea{\end{eqnarray}}
\usepackage{graphicx}

\catcode`\@=11
\def\lsim{\mathrel{\mathpalette\@versim<}}
\def\gsim{\mathrel{\mathpalette\@versim>}}
\def\@versim#1#2{\vcenter{\offinterlineskip
\ialign{$\m@th#1\hfil##\hfil$\crcr#2\crcr\sim\crcr } }}
\catcode`\@=12
\usepackage{axodraw}

\catcode`@=12
\begin{document}
\thispagestyle{empty}
\begin{flushright}
UCRHEP-T464\\
April 2009\
\end{flushright}
\vspace{0.8in}
\begin{center}
{\Large \bf Variants of the Dark Left-Right Gauge Model:\\
Neutrinos and Scotinos\\}
\vspace{0.8in}
{\bf Ernest Ma\\}
\vspace{0.2in}
{\sl Department of Physics and Astronomy, University of California,\\
Riverside, California 92521, USA\\}
\end{center}
\vspace{0.8in}
\begin{abstract}\
In the recently proposed dark left-right gauge model (DLRM) of particle 
interactions, the usual left-handed lepton doublet $(\nu,e)_L$ transforming 
under $SU(2)_L$ is accompanied by the $unusual$ right-handed fermion doublet 
$(n,e)_R$ transforming under $SU(2)_R$, where $n_R$ is $not$ the Dirac mass 
partner of $\nu_L$. In this scenario, whereas $\nu_L$ is certainly a neutrino, 
$n_R$ should be considered a $scotino$, i.e. a dark-matter fermion.  Variants 
of this basic idea are discussed, including its minimal $scotogenic$ 
realization.
\end{abstract}

\newpage
\baselineskip 24pt

\noindent \underline{\it Introduction}~:~ The gauge group $SU(3)_C \times 
SU(2)_L \times U(1)_Y$ of the Standard Model (SM) of particle interactions 
treats left-handed and right-handed fermions differently, with the electric 
charge given by $Q = T_{3L} + Y$.  To restore left-right symmetry, it is 
often proposed that the extension $SU(3)_C \times SU(2)_L \times SU(2)_R 
\times U(1)_{B-L}$ be considered, where $Q = T_{3L} + T_{3R} + (B-L)/2$.  
In that case, the fermion content of the SM gains one extra particle, 
i.e. $\nu_R$ in the right-handed lepton doublet $(\nu,l)_R$.  Connecting 
this with the usual left-handed lepton doublet $(\nu,l)_L$ through a 
Higgs bidoublet, $\nu_R$ pairs with $\nu_L$ to obtain a Dirac mass, 
just as $l_R$ does with $l_L$.  Assuming $SU(2)_R \times U(1)_{B-L}$ is broken 
to $U(1)_Y$ through a Higgs triplet transforming as $(1,1,3,1)$, $\nu_R$ 
gets a large Majorana mass, thereby inducing a small seesaw mass 
for $\nu_L$.  The above is a well-known scenario for what the addition 
of $\nu_R$ would do for understanding the existence of tiny neutrino masses.
For a more general discussion of the $SU(2)_R$ breaking scale, see 
Ref.~\cite{m04}.

Suppose the mass connection between $\nu_R$ and $\nu_L$ is severed without 
affecting $l_R$ and $l_L$, then $\nu_L$ and $\nu_R$ can be different 
particles, with their own interactions.  Whereas $\nu_L$ is clearly still 
the well-known neutrino, $\nu_R$ may become something else entirely. 
As shown in Ref.~\cite{klm09}, it may in fact be a $scotino$, i.e. a 
dark-matter fermion, and to avoid confusion, it is renamed $n_R$. 
This is accomplished in a nonsupersymmetric $SU(3)_C \times SU(2)_L \times 
SU(2)_R \times U(1)$ model with the imposition of a global U(1) symmetry 
$S$, such that the breaking of $SU(2)_R \times S$ will leave the generalized 
lepton number $L = S - T_{3R}$ unbroken.  It is called the dark left-right 
model (DLRM), to distinguish it from the alternative left-right model (ALRM) 
proposed 22 years ago \cite{m87-1,bhm87} which has the same crucial property 
that $n_R$ is $not$ the mass partner of $\nu_L$.

\noindent \underline{\it Fermion content}~:~ The fermion 
structure of the DLRM under $SU(3)_C \times SU(2)_L \times 
SU(2)_R \times U(1) \times S$ is given by \cite{klm09}
\begin{eqnarray}
&& \psi_L = \pmatrix{\nu \cr e}_L \sim (1,2,1,-1/2;1), ~~~ 
\psi_R = \pmatrix{n \cr e}_R \sim (1,1,2,-1/2;1/2), \\   
&& Q_L = \pmatrix{u \cr d}_L \sim (3,2,1,1/6;0), ~~~ d_R \sim (3,1,1,-1/3;0), 
\\ 
&& Q_R = \pmatrix{u \cr h}_R \sim (3,1,2,1/6;1/2), ~~~ h_L \sim (3,1,1,-1/3;1), 
\end{eqnarray}
where $h$ is a new heavy quark of charge $-1/3$.  The above fermionic content 
was first studied in Ref.~\cite{rr78} and also in Ref.~\cite{bm88}, without 
the identification of $n_R$ as a scotino.

To allow $e_L$ to pair with $e_R$ to form a Dirac fermion, the Higgs 
bidoublet
\begin{equation}
\Phi = \pmatrix{\phi_1^0 & \phi_2^+ \cr \phi_1^- & \phi_2^0} \sim (1,2,2,0;1/2)
\end{equation}
is added so that $m_e$ is obtained from $v_2 = \langle \phi_2^0 \rangle$.  
At the same time, $\nu_L$ is connected to $n_R$ through $\phi_1^0$.  However, 
$\langle \phi_1^0 \rangle = 0$ will be maintained because $\phi_1^0$ has 
$S - T_{3R} = 1$, whereas that of $\phi_2^0$ is zero.  As shown in 
Ref.~\cite{klm09}, the spontaneous breaking of $SU(2)_R \times S$ leaves the 
residual symmetry $L = S - T_{3R}$ unbroken, where $L$ is the conventional 
lepton number assigned to $\nu$ and $e$.  Here $n$ has $L = S - T_{3R} = 0$, 
whereas $W_R^\pm$ has $L = S - T_{3R} = \mp 1$, and it does not mix with 
$W_L^\pm$, in contrast to the case of the conventional left-right model, 
where such mixing is unavoidable.  Further, the bidoublet
\begin{equation}
\tilde{\Phi} = \sigma_2 \Phi^* \sigma_2 = \pmatrix{\bar{\phi}_2^0 & -\phi_1^+ 
\cr -\phi_2^- & \bar{\phi}_1^0} \sim (1,2,2,0;-1/2)
\end{equation}
is prevented by $S$ from coupling $\psi_L$ to $\psi_R$, thereby ensuring the 
absence of tree-level flavor-changing neutral currents, which was not 
possible in the conventional nonsupersymmetric left-right model.

In the quark sector, $Q_L$ couples to $Q_R$ through $\tilde{\Phi}$, but not 
$\Phi$.  Hence $m_u$ is obtained from $v_2$, and there is no mixing between 
$d$ and $h$. The former has $L=0$, but the latter has $L = S - T_{3R} = 1$.  
For $d_L$ to pair with $d_R$, and $h_R$ to pair with $h_L$, the Higgs doublets
\begin{equation}
\Phi_L = \pmatrix{\phi_L^+ \cr \phi_L^0} \sim (1,2,1,1/2;0), ~~~ 
\Phi_R = \pmatrix{\phi_R^+ \cr \phi_R^0} \sim (1,1,2,1/2;-1/2)
\end{equation}
are needed.  Note that $v_4 = \langle \phi_R^0 \rangle$ will break 
$SU(2)_R \times U(1)$ to $U(1)_Y$ as desired, and the leptoquark $h$  
gets a heavy mass of order $v_4$.

\noindent \underline{\it Exotic variants}~:~ The fermion sector may be 
more exotic.  For example, Eq.~(1) may be replaced by
\begin{eqnarray}
&& \psi_L = \pmatrix{\nu \cr e}_L \sim (1,2,1,-1/2;1), ~~~ 
e_R \sim (1,1,1,-1;1),\\
&& \psi_R = \pmatrix{n \cr E}_R \sim (1,1,2,-1/2;-1/2), ~~~ 
E_L \sim (1,1,1,-1;0).
\end{eqnarray}
In this case, $E$ has $L=0$, $n$ has $L=-1$, $m_e$ comes from $v_3 = \langle 
\phi_L^0 \rangle$, $m_E$ from $v_4$, and neither $\Phi$ nor $\tilde{\Phi}$ 
couples to $\bar{\psi}_L \psi_R$.

As another example, Eqs.~(2) and (3) may be replaced by
\begin{eqnarray}
&& Q_L = \pmatrix{u \cr d}_L \sim (3,2,1,1/6;0), ~~~ u_R \sim (3,1,1,2/3;0), \\
&& Q_R = \pmatrix{f \cr d}_L \sim (3,1,2,1/6;-1/2), ~~~ 
f_L \sim (3,1,1,2/3;-1).
\end{eqnarray}
In this case, $f$ is an exotic quark of charge 2/3 and $L=-1$.  Here, $Q_L$ 
couples to $Q_R$ through $\Phi$, but not $\tilde{\Phi}$, $m_d$ comes from 
$v_2$, $m_u$ from $v_3$, and $m_f$ from $v_4$.

\noindent \underline{\it Masses for $\nu_L$ and $n_R$}~:~ With the above 
Higgs content, $\nu_L$ and $n_R$ remain massless.  Consider now the various 
ways that they acquire masses:

\noindent (1) In the DLRM \cite{klm09}, Higgs triplets under $SU(2)_L$ 
and $SU(2)_R$ are used separately for $\nu_L$ and $n_R$ masses.  

\noindent (2) In Ref.~\cite{rr78}, they are massless.

\noindent (3) In Ref.~\cite{bm88}, they acquire radiative masses separately 
from the addition of two charged scalar singlets.

\noindent (4) In the ALRM \cite{m87-1}, the usual lepton doublet is actually 
part of a bidoublet:
\begin{equation}
\pmatrix{\nu_e & E^c \cr e & N^c_E}_L \sim (1,2,2,0),
\end{equation}
which means that $\nu_e$ and $e$ have $SU(2)_R$ interactions. In the 
original proposal, $\nu_L$ and $n_R$ are massless, but they can acquire 
seesaw masses separately through the many other fields available in the 
\underline{27} representation of $E_6$, as explained in Ref.~\cite{m00}. 
One of the three $n_R$ copies in this supersymmetric model pairs with 
the neutral gaugino from the breaking of $SU(2)_R \times U(1) \to U(1)_Y$ 
to form a Dirac fermion.  The other two are light and considered as 
sterile neutrinos which mix with $\nu_L$ through the soft term 
$\nu_e N^c_E - e E^c$ which breaks $R$ parity.

\noindent (5) A simple variation of the DLRM also allows neutrino masses to 
be radiatively generated by dark matter (i.e. {\it scotogenic}) in one loop 
\cite{m06-1,m06-2,m06-3,kms06,ks06,kko07,hkmr07,bm08,m08-1,m08-2,m08-3,
m08-4,akrsz09,ms09,m09,sty09}. Instead of $\Delta_L$, a scalar singlet 
$\chi \sim (1,1,1,0;-1)$ is added, then the trilinear scalar term $Tr(\Phi 
\tilde{\Phi}^\dagger) \chi$ is allowed. 
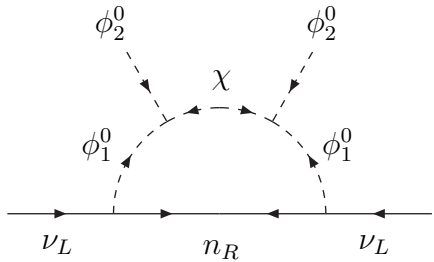
\begin{figure}[htb]
\begin{center}
\begin{picture}(500,100)(120,45)
\ArrowLine(270,50)(310,50)
\ArrowLine(390,50)(350,50)
\ArrowLine(310,50)(350,50)
\ArrowLine(430,50)(390,50)
\Text(290,35)[b]{$\nu_L$}
\Text(410,35)[b]{$\nu_L$}
\Text(352,33)[b]{$n_R$}
\Text(352,97)[b]{$\chi$}
\Text(305,70)[b]{$\phi^0_1$}
\Text(396,70)[b]{$\phi^0_1$}
\Text(310,116)[b]{$\phi^0_2$}
\Text(390,116)[b]{$\phi^0_2$}
\DashArrowLine(315,111)(330,85){3}
\DashArrowLine(385,111)(370,85){3}
\DashArrowArcn(350,50)(40,180,120){3}
\DashArrowArc(350,50)(40,0,60){3}
\DashArrowArc(350,50)(40,90,120){3}
\DashArrowArcn(350,50)(40,90,60){3}
\end{picture}
\end{center}
\caption[]{One-loop scotogenic neutrino mass.}
\end{figure}
Using the soft term $\chi^2$ to break $L$ to $(-)^L$, a scotogenic neutrino 
mass is obtained as shown in Fig.~1.  It is also possible to do this in two 
loops \cite{ms07,m08-5} and three loops \cite{knt03,cs04,aks09}.

\noindent (6) Since the scotogenic mechanism of Fig.~1 does not care how $n_R$ 
acquires a Majorana mass, it may be accomplished with three neutral singlet 
fermions $n_L$ with $S=0$ instead of the Higgs triplet $\Delta_R$.  Now 
the Yukawa coupling $\bar{n}_L (n_R \phi^0_R - e_R \phi_R^+)$ is allowed, 
as well as a Majorana mass for $n_L$.  Hence $n_R$ gets an induced Majorana 
mass which is essential for Fig.~1.  Note that $n_L$ does not couple to 
$(\nu_L \phi_L^0 - e_L \phi_L^+)$ because of $S$.

In this minimal variant, the $Z'$ mass comes entirely from the $\Phi_R$ doublet 
as in the ALRM, hence the prediction
\begin{equation}
M^2_{W_R} = {(1-2x) \over (1-x)} M^2_{Z'} + {x^2 \over (1-x)^2} M^2_{W_L},
\end{equation}
where $x \equiv \sin^2 \theta_W$ and zero $Z-Z'$ mixing has been assumed 
\cite{bhm87}, i.e. $v_2^2/(v_2^2+v_3^2) = x/(1-x)$.  Currently, the 
experimental bound on $M_{Z'}$ is 850 GeV \cite{klm09}.

The diagram of Fig.~1 is exactly calculable \cite{m06-1}. The $\bar{n}_R \nu_L 
\bar{\phi}_1^0$ coupling is given by $(m_\alpha/v_2) U_{\alpha i}$, where 
$\alpha = e,\mu,\tau$ and $i = 1$ to 6 refer to the mass eigenstates of the 
$6 \times 6$ $(n_R,n_L)$ mass matrix
\begin{equation}
{\cal M}_n = \pmatrix{0 & m_D \cr m_D & m_n}.
\end{equation}
In the bases (Re$\phi_1^0$, Re$\chi$), (Im$\phi_1^0$, Im$\chi$), the respective 
mass-squared matrices are
\begin{equation}
\pmatrix{m_\phi^2 & \mu v_2 \cr \mu v_2 & m_\chi^2 + \mu_\chi^2}, ~~~ 
\pmatrix{m_\phi^2 & -\mu v_2 \cr -\mu v_2 & m_\chi^2 - \mu_\chi^2}.
\end{equation}
Let their mass eigenstates and mixing angles be $(m^2_{R1}, m^2_{R2}, \theta_R)$ 
and $(m^2_{I1}, m^2_{I2}, \theta_I)$, then
\begin{eqnarray}
({\cal M}_\nu)_{\alpha \beta} &=& \sum_i {m_\alpha m_\beta U_{\alpha_i} U_{\beta i} 
M_i  \over 16 \pi^2 v_2^2} ~[~ \cos^2 \theta_R {m^2_{R1} \over m^2_{R1} - M_i^2} 
\ln {m^2_{R1} \over M_i^2} - \cos^2 \theta_I {m^2_{I1} \over m^2_{I1} - M_i^2} 
\ln {m^2_{I1} \over M_i^2} \nonumber \\ && + \sin^2 \theta_R {m^2_{R2} \over 
m^2_{R2} - M_i^2} \ln {m^2_{R2} \over M_i^2} - \sin^2 \theta_I {m^2_{I2} \over 
m^2_{I2} - M_i^2} \ln {m^2_{I2} \over M_i^2}~].
\end{eqnarray}
In the limit $\mu_\chi^2 = 0$, ${\cal M}_\nu$ vanishes because $m_{R1} = m_{I1}$,
$m_{R2} = m_{I2}$, and $\theta_R = -\theta_I$.  In the limit $\mu = 0$, 
${\cal M}_\nu$ also vanishes because $m_{R1} = m_{I1}$ and $\theta_R = \theta_I 
= 0$.  Furthermore, if $m_n = 0$ in Eq.~(13), ${\cal M}_\nu$ is again zero 
because $n$ is a Dirac particle.  This latter is an example of the inverse 
seesaw mechanism \cite{ww83,mv86,m87-2,dv05,kk07}.  Hence neutrino masses are 
suppressed in this scenario by three possible limits, and the scale of 
$SU(2)_R$ breaking (associated with the masses of the dark-matter particles 
of Fig.~1) may well be as low as 1 TeV, as advocated.

\noindent \underline{\it Conclusion}~:~ The neutral component $n$ of the 
$SU(2)_R$ doublet $(n,e)_R$ is proposed as a dark-matter fermion (scotino). 
Variants of this basic idea, the dark left-right model (DLRM), are discussed. 
A minimal version is considered, where neutrino masses are radiatively 
generated by dark matter (scotogenic) and naturally suppressed, allowing 
the $SU(2)_R$ breaking scale to be as low as 1 TeV.

\noindent \underline{\it Acknowledgment}~:~ This work was supported in part 
by the U.~S.~Department of Energy under Grant No.~DE-FG03-94ER40837.

\newpage
\bibliographystyle{unsrt}

\end{document}